\documentclass[pre, onecolumn]{revtex4}
\usepackage{amsmath}
\usepackage{amssymb}
\usepackage{amscd}
\usepackage{graphicx}
\usepackage{hyperref}

\begin{document}
\title{Global Defect Topology in Nematic Liquid Crystals}
\author{Thomas Machon}
\affiliation{Department of Physics and Centre for Complexity Science, University of Warwick, Coventry, CV4 7AL, United Kingdom.}
\author{Gareth P. Alexander}
\affiliation{Department of Physics and Centre for Complexity Science, University of Warwick, Coventry, CV4 7AL, United Kingdom.}

\begin{abstract}
We give the global homotopy classification of nematic textures for a general domain with weak anchoring boundary conditions and arbitrary defect set in terms of twisted cohomology, and give an explicit computation for the case of knotted and linked defects in $\mathbb{R}^3$, showing that the distinct homotopy classes have a 1-1 correspondence with the first homology group of the branched double cover, branched over the disclination loops. We show further that the subset of those classes corresponding to elements of order 2 in this group have representatives that are planar and characterise the obstruction for other classes in terms of merons. The planar textures are a feature of the global defect topology that is not reflected in any local characterisation. Finally, we describe how the global classification relates to recent experiments on nematic droplets and how elements of order 4 relate to the presence of $\tau$ lines in cholesterics.
\end{abstract}
\maketitle


The textures of liquid crystals and other ordered media have a long association with topology. This arose initially through Frank's introduction of the term {\sl disinclination} (now disclination) for the characteristic lines, or threads, that give nematics their name, together with a half-integer winding number to classify them~\cite{frank58}. In the 1970s the homotopy theory of defects was developed, providing a description of discontinuities in ordered media in terms of conjugacy classes of the homotopy groups $\pi_n(G/H)$, where $G$ is the symmetry group of the high temperature disordered phase and $H$ the unbroken isotropy subgroup of the ordered phase~\cite{mermin79}. For nematic liquid crystals $G=SO(3)$ may be taken to be the rotational symmetry group of Euclidean $\mathbb{R}^3$ and $H=D_{\infty}$ the subgroup corresponding to the symmetries of a cylinder, or rod. The ground state manifold $G/H\cong\mathbb{RP}^2$ is the real projective plane and it follows that the line defects, or disclinations, correspond to elements of $\pi_1(\mathbb{RP}^2)\cong\mathbb{Z}_2$ and the point defects to the conjugacy classes of $\pi_2(\mathbb{RP}^2)\cong\mathbb{Z}$. These latter are simply the pairs $(q,-q)$, so that nematic point defects are classified by $\mathbb{Z}/(q\sim-q)\cong\mathbb{N}$. 

Modern experiments allow for the controlled creation of defects and their manipulation to produce textures that may serve as soft photonic elements~\cite{musevic13,ackerman12}, novel metamaterials~\cite{musevic06,wood11,ravnik11,lavrentovich11,nych13}, or topologically stabilised memory devices~\cite{araki11,musevic11}. 
In many cases this is fascilitated by the immersion of colloidal particles in the liquid crystal, whose surface anchoring properties are used to imprint features on the nematic director. Arrays of spherical particles allow defect lines that entangle them to be controllably reconfigured so as to form any knot or link~\cite{tkalec11,jampani11}, providing a practical realisation of aspects of knot theory in soft materials~\cite{copar15}. Such particles can also be dressed by Skyrmion-like excitations that modify and augment the topology~\cite{pandey14}. 
In addition, the topology of the colloids themselves can be controlled to produce handlebodies~\cite{senyuk13,liu13,cavallaro13}, M\"obius strips~\cite{machon13}, knots~\cite{martinez14} and even linked particles~\cite{martinez15}. In these systems it is the combined properties of the defects they engender in the nematic order, and the global textures they create, that are of principal interest rather than the localised characterisation of individual defects that is the focus of the traditional homotopy theory. Thus in characterising them one would like to determine the global topology of the entire nematic texture and how it relates to the domain. A feature here is that the region occupied by the liquid crystal, or more correctly where the director is well-defined, {\sl i.e.} the region exterior to any inclusions and also excluding the defects, is itself a non-trivial space. For a general set of disclinations it is the complement of a link and the topological properties of the liquid crystal will depend on, and reflect, the topology of this domain. It is clear that the topological classification of nematic textures in such a domain will correspond to a link invariant: in \cite{machon14} we stated that this is the first homology group of the double branched cover of the link complement. This group is presented by the Goeritz matrix or the Gordon-Litherland form~\cite{Lickorish,gordon78} both of which are readily computable from an ordinary knot diagram, fascilitating explicit calculation of the homotopy classification for any case of interest. For instance, from this one finds that there are 8 (pointed) homotopy classes of textures on the complement of the Whitehead link, but only a single class on the complement of the Kinoshita-Terasaka knot. Here we describe the global topology of defects in nematics for a general domain, including a detailed account of the classification of knotted disclinations, and develop a number of extensions and explicit worked examples of direct relevance to currently realisable experiments and aimed at illustrating the richness that is present in the global topology of nematics. 

An introductory example illustrating the general classification problem we describe is provided by a pair of linked disclination loops forming the Hopf link, for which there are two homotopically distinct nematic textures. The most direct way to see this is to observe that the complement of the Hopf link in $S^3$ has the homotopy type of a torus. Thus the classification is the same as that of maps $T^2\to\mathbb{RP}^2$ with the property that both the meridian and longitude of the torus are sent to the non-trivial element of $\pi_1(\mathbb{RP}^2)$, since each goes around one of the two disclination lines making up the Hopf link. Such maps have been classified by J\"anich~\cite{janich87}, who showed that there are only two. The result has been revisited several times in light of recent experimental advances. A geometric construction developed by \v{C}opar and \v{Z}umer~\cite{copar11} tracks the local profile of each disclination loop along their contour length to define a self-linking number, making direct use of the nature of the textures observed around defect lines in experiments with spherical colloids. This self-linking number is shown to be invariant modulo two. A complementary approach views the dichotomy in terms of a `hedgehog' or `Skyrmion' charge, analogous to the element of $\pi_2(\mathbb{RP}^2)$ that classifies point defects, with the subtlety that the non-orientability around the cycles of the torus allows it to be reduced modulo two~\cite{alexander12,BryanThesis}. The two Hopf link textures can be distinguished in experimental realisations by looking at images taken under crossed polarisers; lighter (and darker) areas of the image form a surface whose boundary is the link, and the two homotopy classes differ in the linking number induced by this surface~\cite{copar15,machon14}. 
This example of the Hopf link captures the general structure of the classification for any link. The material domain has the homotopy type of a 2-complex, $X$, so that the classification is given by maps $X\to\mathbb{RP}^2$ with the property that around each disclination line they induce the non-trivial element of $\pi_1(\mathbb{RP}^2)$. A second type of data, the analogues of degrees or elements of $\pi_2(\mathbb{RP}^2)$, can be associated to each two-dimensional cell of the complex, which, as for the Hopf link, is subject to a reduction coming from the non-orientable behaviour of the director around the 1-skeleton of the complex. This is an example of a twisted cocycle, representing a cohomology class with local coefficients. A general strategy for computing such invariants using obstruction theory was provided by Steenrod in his classic text on fibre bundles~\cite{Steenrod} and is essentially the approach we employ here. 

The pointed homotopy classes of nematic textures are in one-to-one correspondence with the elements of an Abelian group, the first homology group of the double branched cover, branched over the disclination loops. However, there is, in general, no canonical way to identify the set of homotopy classes with the group structure. For instance, in the example of the Hopf link both textures could equally be regarded as the $0$ element of $\mathbb{Z}_2$. Nonetheless, some elements of the group structure do correspond to physical properties of the liquid crystal texture. Again, in the example of the Hopf link this is that both homotopy classes have representatives as planar textures, {\sl i.e.} in which the director lies everywhere in the $xy$-plane, having no component along $z$. This is not a property exhibited by every homotopy class in the general case; we show here that it is only true for those homotopy classes that correspond to elements of order $2$. Thus, for example, in the case of the Borromean rings where the group is $\mathbb{Z}_4\oplus\mathbb{Z}_4$ there are $16$ homotopy classes of nematic textures~\cite{note1}, $4$ of which have representatives that are planar. Part of the physical significance of such planar textures may be conveyed as follows. Liquid crystals reorient readily in response to applied electric or magnetic fields. For materials with negative dielectric (or diamagnetic) anisotropy the director reorients to lie orthogonal to the direction of the applied field, so that if the field is applied in the $z$-direction the director will lie primarily, and if possible exclusively, in the $xy$-plane. The homotopy classes with planar representatives correspond to the low energy states when a material with negative dielectric anisotropy is placed in a uniform electric field. In these conditions, the homotopy classes that do not possess planar representatives will necessarily exhibit regions where the director is parallel (or anti-parallel) to the field, though for energetic reasons such regions will be spatially localised, like the familiar $\pi$-walls~\cite{deGennesProst}. These regions of localised non-planarity are merons that can be thought of as a fractionalisation of Skyrmions. Where they cannot be eliminated to give a purely planar texture they represent (in the sense of Poincar\'e duality) an obstruction and so reflect the non-trivial topology of the texture. 

The global approach to defect topology provides a classification of nematic textures in terms of topological properties of the entire domain and can be contrasted against the homotopy groups which reflect only the behaviour of the director field in the immediate vicinity of each defect. It reveals an essential dichotomy between homotopy classes of textures on the complement of knotted and linked disclination loops: that of planar and non-planar textures. While any homotopy class can be related to any other through decoration with Skyrmion-like distortions, in some cases they can be removed by homotopy -- the planar textures -- and in some cases they cannot. This difference, as well as the more general homotopy classification, is fundamentally global in nature and is not captured by local data on the boundary of each link component. Algebraically, the planar textures associated to the link are enumerated by the order 2 subgroup of the first homology of the branched double cover. Geometrically they are more subtle, and can be associated to distinct spanning surfaces for the link. In the case of cholesterics the Skyrmion-like distortions that distinguish non-planar textures are realised as $\lambda$ lines (defects in the pitch) and the global theory gives constraints on their total number.

\section{Homotopy Classification of Nematic Textures}

To study defects from a global perspective we consider homotopy classes of maps from a material domain $\Omega$ into $\mathbb{R} \mathbb{P}^2$. The domain may incorporate inclusions such as colloidal particles as well as defects so that, in general, $\Omega$ will be an open subset of $\mathbb{R}^3$ from which a neighbourhood of a collection of points and a collection of closed loops, representing both colloidal inclusions and defects in the nematic director, has been removed. For example, for a single point defect at the origin, the domain is $\mathbb{R}^3-N(0)$ and for a knotted or linked defect, $L$, the domain is $\mathbb{R}^3-N(L)$, where $N$ denotes an open neighbourhood. We impose free boundary conditions (up to a topological class) on all defects and other boundaries (such as colloidal particles) in the system. This corresponds to weak anchoring conditions on the colloidal inclusions. As a consequence of this, our classification will not consider any phenomena associated to the Hopf invariant, $\pi_3(\mathbb{RP}^2)=\mathbb{Z}$, though we would expect that it may play a role in cases with fixed boundary conditions and/or the case of a periodic domain. 

By our assumptions $\Omega$ will be homotopy equivalent to a 2-complex~\cite{Lickorish} and the problem becomes homotopy classes of maps from an orientable 2-complex, $X(\Omega)$, into $\mathbb{RP}^2$. We note that this also includes closed surfaces as domains, for example a torus enclosing a defect line, which will serve as a simple example in the text. The homotopy classification of maps from a general $p$-complex into $p$-projective space has been solved through obstruction theory by Olum~\cite{olum62}, with accounts for the case of surfaces given by Eells and Lemaire~\cite{eells80}, as well as by Adams~\cite{adams82}. The result is that free homotopy classes of maps from an oriented 2-complex are given by two invariants, associated to $\pi_1(\mathbb{RP}^2)$ and $\pi_2(\mathbb{RP}^2)$ respectively. 

\subsection{First Invariant}
The first invariant of the texture describes its orientability. Associated to each closed loop, $\gamma$, in the domain is an element of $\pi_1(\mathbb{R} \mathbb{P}^2) = \mathbb{Z}_2$ which records whether the nematic preserves or reverses orientation around $\gamma$. One thus obtains a homomorphism of fundamental groups
\begin{equation}
\theta : \pi_1(\Omega) \to \pi_1 (\mathbb{R} \mathbb{P}^2).
\end{equation}
$\theta$ factors through the Abelianisation of $\pi_1(\Omega)$ to a map on the first homology $H_1(\Omega) \to \mathbb{Z}_2$ and thus defines a cocyle~\cite{note2} 
\begin{equation}
w_1(\mathbf{n}) \in H^1(\Omega ; \mathbb{Z}_2),
\end{equation}
which we will refer to as the first invariant for a nematic texture. From a physical perspective, $w_1(\mathbf{n})$ can be thought of as a $\mathbb{Z}_2$ gauge field~\cite{lammert93,lammert95}. As an example, for a nematic in $\mathbb{R}^3$ containing $N$ disclinations, $H^1(\Omega ; \mathbb{Z}_2) = \mathbb{Z}_2^N$. Since, by definition, $\mathbf{n}$ is nonorientable around disclination lines, in this case $w_1(\mathbf{n}) = (1, 1,\ldots,1) \in \mathbb{Z}_2^N$. If, instead, the first of these lines corresponds to a toroidal colloidal particle~\cite{senyuk13}, or another non-trivial structure in the material domain, around which the liquid crystal is orientable, then $w_1(\mathbf{n}) = (0,1,\ldots,1) \in \mathbb{Z}_2^N$. Note that the global information about the orientability of the nematic is just the sum of the local information around each loop.

\subsection{Skyrmion Data}
The second type of topological data characterising a nematic texture is associated to Skyrmions: to each surface in the domain, closed or with boundary contained in the boundary of the domain, one can assign an element of $\pi_2(\mathbb{RP}^2)=\mathbb{Z}$, which records the number of Skyrmions on that surface. Thus we obtain a cocycle in degree two. The Skyrmions can be moved around the sample under homotopy and in doing so acquire a twisting coming from the non-trivial action of $\pi_1(\mathbb{RP}^2)$ on $\pi_2(\mathbb{RP}^2)$. The general algebraic statement of this is that the topologically distinct ways of adding Skyrmion-like distortions to a nematic texture are in one-to-one correspondence with elements in the set
\begin{equation}
H^2(\Omega ; \mathbb{Z}^{w_1}) / (x \sim -x),
\label{eq:second}
\end{equation}
where $H^2(\Omega ; \mathbb{Z}^{w_1})$ is the twisted cohomology group with the local coefficient system $\mathbb{Z}^{w_1}$ given by the group of integers along with the homomorphism $\theta: \pi_1(\Omega) \to \mathbb{Z}_2$, where $\mathbb{Z}_2$ is now thought of as the automorphism group of $\mathbb{Z}$ and acts through multiplication by $-1$ around a non-orientable loop, describing the reversal of Skyrmion charge under the antipodal map. The global equivalence $x \sim -x$ accounts for the passage from pointed to free homotopy classes of maps, the generalisation of the well-known $\mathbb{Z} \to \mathbb{N}$ reduction in the case of a single point defect.

In general the particular element of $H^2(\Omega ; \mathbb{Z}^{w_1})$ that a given pointed homotopy class corresponds to cannot be identified; as we will show in \S\ref{sec:planar} all elements of order $2$ serve as equivalent choices for the null map. However, by choosing a reference map on the 1-skeleton of $X(\Omega)$, (as is done, in~\cite{olum62} and~\cite{eells80}), one may identify pointed homotopy classes of textures with elements in $H^2(\Omega ; \mathbb{Z}^{w_1})$.

\section{Computation}

To give an effective computation of the group $H^2(\Omega ; \mathbb{Z}^{w_1})$ we will make use of a double cover for the domain $\Omega$, denoted $\Omega^{w_1}$, defined~\cite{note3} so that $\mathbf{n}$ is orientable along the projection of any loop in $\Omega^{w_1}$. By construction, one can then lift the director field on $\Omega$ to an orientable field on the double cover to create a map $\hat{\mathbf{n}}: \Omega^{w_1} \to S^2$. Associated to $\Omega^{w_1}$ is the deck transformation $t$, $t^2=1$, which permutes the sheets of the cover, and consistency of the lift demands that $\hat{\mathbf{n}}$ reverses orientation on switching between corresponding points on the two sheets, $\Omega_1$ and $\Omega_2$, of the cover. This can be phrased as the $\mathbb{Z}_2$ equivariance condition $\hat{\mathbf{n}}(x) = - \hat{\mathbf{n}}(t x)$, or equivalently demanding that the following diagram commutes
\begin{equation}
\begin{CD}
\Omega^{w_1} @ > \hat{\mathbf{n}} >>S^2\\
@V p_1VV @V p_2 VV \\
\Omega @ > \mathbf{n} >> \mathbb{R} \mathbb{P}^2
\end{CD}
\end{equation}
where $p_1$ and $p_2$ are the projection maps associated to the two covering spaces $\Omega^{w_1}$ and $S^2$. Homotopy classes of nematic textures can then be classified by $\mathbb{Z}_2$ equivariant maps $\Omega^{w_1} \to S^2$, denoted $[\Omega^{w_1}, S^2]^\ast_{eq}$. Arbitrary maps, not necessarily equivariant, are classified by the second cohomology group, $H^2(\Omega^{w_1}; \mathbb{Z})$, and the equivariant maps are correspondingly classified by the equivariant cohomology~\cite{note4} $H^2(\Omega^{w_1}; \mathbb{Z})_{eq}$, equivalently the twisted cohomology of $\Omega$. Taking into account the ambiguity of sign in the lift from $\mathbf{n}$ to $\hat{\mathbf{n}}$ gives an equivalence relation between $x$ and $-x$ in $H^2(\Omega; \mathbb{Z}^{w_1})$.

\begin{figure}
\begin{center}
\includegraphics{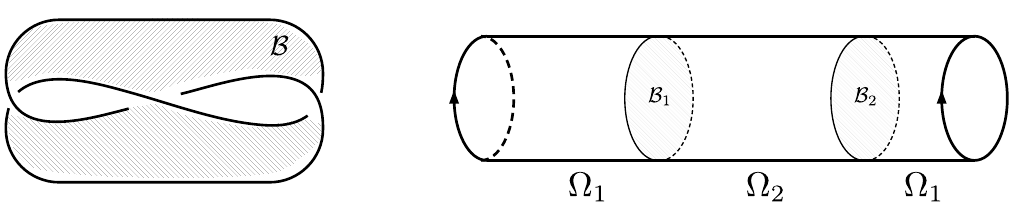}
\caption{\textit{Left}: Branch set for a trefoil knot disclination line, equivalent to a Seifert surface. \textit{Right}: Schematic depiction of the covering space $\Omega^{w_1}$, the two sheets, $\Omega_1$ and $\Omega_2$ are glued together along two copies of the branch set.}
\label{fig:branch}
\end{center}
\end{figure}

For computational purposes it is useful to describe the covering space $\Omega^{w_1}$ in terms of branch sets, illustrated in Figure \ref{fig:branch}, whose topology is determined by the Poincar\'{e} dual of $w_1(\mathbf{n})$, $PD[w_1] \in H_{2}(\Omega, \partial \Omega; \mathbb{Z}_2)$. For our purposes a branch set $\mathcal{B}$ will be a connected properly embedded orientable~\cite{note5} codimension 1 submanifold of $\Omega$ such that $[\mathcal{B}] = PD[w_1(\mathbf{n})]$. The geometric interpretation of $\mathcal{B}$ is an orientable submanifold of $\Omega$ chosen such that if $\gamma$ is an oriented loop in $\Omega$, then the intersection number of $\gamma$ and $\mathcal{B}$ is even if $\mathbf{n}$ is orientable around $\gamma$ and odd if $\mathbf{n}$ is nonorientable around $\gamma$, or
\begin{equation}
\textrm{Int}(\mathcal{B}, \gamma ) \equiv \theta(\gamma) \quad \textrm{mod} \; 2.
\end{equation}

An example of relevance is the case $\Omega = \mathbb{R}^3-N(L)$, where $N(L)$ is a neighbourhood of a set of disclination lines $L$. Since $\mathbf{n}$ is non-orientable around the line defects, $\mathcal{B}$ forms a Seifert surface for $L$, illustrated in Figure \ref{fig:branch}. In terms of this branch set, the double cover $\Omega^{w_1}$ is formed by taking two copies of $\Omega$, $\Omega_1$ and $\Omega_2$, each cut along $\mathcal{B}$, and then gluing the pieces together so that the top side of $\mathcal{B}$ in $\Omega_1$ is glued to the bottom side of $\mathcal{B}$ in $\Omega_2$ and vice versa, illustrated in Figure \ref{fig:branch}~\cite{note6}.

To compute $H^2(\Omega ; \mathbb{Z}^{w_1})$ when the map $\theta$ is nontrivial we will pass through Poincar\'{e}-Lefschetz duality and instead compute the isomorphic twisted relative homology group, $H_{1}(\Omega, \partial \Omega ; \mathbb{Z}^{w_1})$, and equivalently the equivariant homology group on the cover $\Omega^{w_1}$. The decomposition of $\Omega^{w_1}$ through the branch set $\mathcal{B}$ allows one to write down a Mayer-Vietoris sequence 
\begin{equation}
\begin{CD}
\to H_n(\mathcal{B}^{w_1}, \partial \mathcal{B}^{w_1})_{eq} @> i>> \left (H_n(\Omega_1, \partial_1) \oplus H_n(\Omega_2, \partial_2) \right )_{eq} @ >>> H_n(\Omega , \partial \Omega ; \mathbb{Z}^{w_1}) \to
\end{CD}
\label{eq:mayer}
\end{equation}
where $\partial_i = \partial \Omega_i \cap  \partial \Omega^{w_1}$. As $\mathcal{B}^{w_1}$ has two components, $\mathcal{B}_1$ and $\mathcal{B}_2$, the homology splits $H_\bullet(\mathcal{B}, \partial \mathcal{B}) \cong H_\bullet(\mathcal{B}_1, \partial \mathcal{B}_1) \oplus H_\bullet(\mathcal{B}_2, \partial \mathcal{B}_2)$. The equivariant homology of $\mathcal{B}^{w_1}$ then consists of elements of $H_\bullet(\mathcal{B}, \partial \mathcal{B})$ that are of the form $(x, -x)$. Similarly $ \left (H_1(\Omega_1, \partial_1) \oplus H_1(\Omega_2, \partial_2) \right)_{eq}$ consists of elements of the form $(y,-y)$. The equivariance condition means that the behaviour in one sheet of the cover determines the behaviour on both. As such, if $H_{0}(\mathcal{B}^{w_1}, \partial \mathcal{B}^{w_1})=0$, as will be the case in our examples, then the twisted cohomology may be computed from information on one sheet as 
\begin{equation}
H^{2}(\Omega;\mathbb{Z}^{w_1}) = H_{1}( \Omega - \mathcal{B}, \partial \Omega) / R,
\label{eq:comp}
\end{equation}
where $R$ are the set of relations determined by the inclusion map $i$ in \eqref{eq:mayer}, restricted to $\Omega_1$. To define the relations $R$, we write $i$ as $i= i_1+ i_2$, corresponding to inclusions from $\mathcal{B}_1$ and $\mathcal{B}_2$ respectively. Giving $\mathcal{B}^{w_1}$ an orientation invariant under the deck transformation we can further split $i$ as $i_1^+ - i_1^- + i_2^+ - i_2^-$, where $\pm$ denotes the inclusion in the positive (negative) direction as defined by the orientations. The restriction to $\Omega_1$ gives $i|_{\Omega_1}=i_2^{+}-i_1^{-}$ and taking into account the equivariant form of \eqref{eq:mayer} allows the relations $R$ to be written as 
\begin{equation}
i^+x + i^- x =0, 
\label{eq:R}
\end{equation}
where $x \in H_{1}(\mathcal{B}, \partial \mathcal{B})$ is now a relative homology cycle in the downstairs branch set.

\subsection{Nematic Textures on the Torus and Other Surfaces}

As a first illustration, we give the computation of homotopy classes of nematic textures on a toroidal domain, $\Omega=T^2$, using the formalism we have just described. This is a classic calculation~\cite{janich87} that is used to model the neighbourhood of a defect line by setting $\mathbf{n}$ to be non-orientable around the meridian of the torus. It is also homotopy equivalent to the complement of the Hopf link in $S^3$ and so models a domain corresponding to the complement of linked defects, by setting ${\bf n}$ to be non-orientable around both the meridian and longitude of the torus. It is easy to see that \eqref{eq:comp} and \eqref{eq:R} continue to apply, except that the relative homology groups are in degree 0 rather than degree 1~\cite{note7}. 

The first step in the classification is to give the possible first invariants, or maps $\pi_1(T^2) \to \mathbb{Z}_2$. The fundamental group of the torus is $\pi_1(T^2) = \mathbb{Z}^2$, generated by the meridian and a longitude, and so the maps are specified by whether $\mathbf{n}$ is orientable or not along each of the meridian and longitude of the torus. If $w_1(\mathbf{n})$ is trivial, then the texture is orientable and one can immediately lift to a map $T^2 \to S^2$, which is classified by degree. Taking into account the relation $x \sim -x$, one finds that the classification of textures is given by the natural numbers, $\mathbb{N}$, as in the case of the sphere. 

\begin{figure}
\begin{center}
\includegraphics{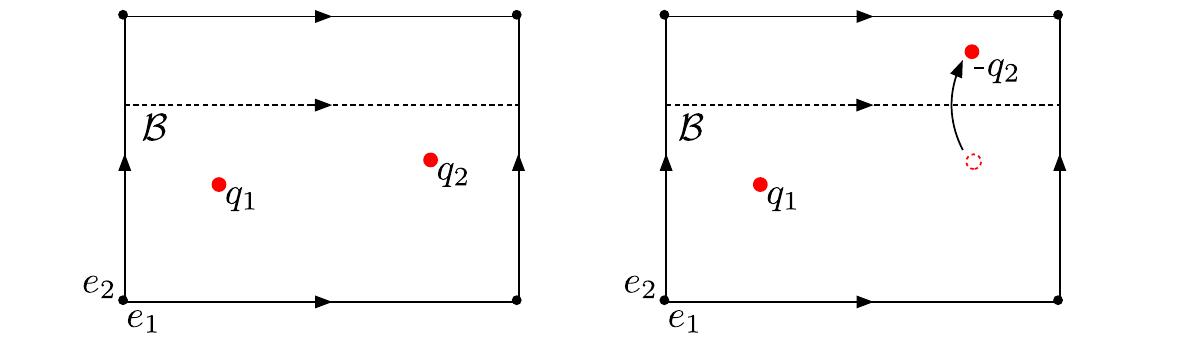}
\caption{Schematic depiction of twisted homology on the torus. $e_1$ and $e_2$ label the longitude and meridian respectively. We assume the nematic is non-orientable along $e_2$, so $\theta(e_1)=0$ and $\theta(e_2)=1$. A branch set $\mathcal{B}$ must intersect $e_2$ an odd number of times, which can be achieved by setting it equal (homologous) to $e_1$. Elements in the twisted homology $H_0(\Omega; \mathbb{Z}^{w_1})$ can then be described by the total charge of a set point charges on the torus whose sign flips upon passing through $\mathcal{B}$. This establishes an equivalence between $q_1+q_2$ and $q_1-q_2$, for any $q_1,q_2 \in \mathbb{Z}$, and so gives the topological classification as $\mathbb{Z}_2$.}
\label{fig:torus_homotopy}
\end{center}
\end{figure}

In the non-orientable case, $w_1({\bf n})$ non-trivial, we need to compute the twisted homology group $H_0(T^2; \mathbb{Z}^{w_1})$. We first construct a branch set $\mathcal{B}$. If $e_1$ and $e_2$ are a longitude and meridian respectively, then the intersection form on the first homology is given in terms of Pauli matrices by $i \sigma_y$, which determines the homology class of possible branch sets. In the example shown in Figure \ref{fig:torus_homotopy}, $\mathbf{n}$ is non-orientable along $e_2$ and so $\mathcal{B}$ is chosen as an embedded circle homologous to $e_1$. In this case we have $H_0(T^2 - \mathcal{B}) = \mathbb{Z}$, the inclusion maps, $i^+$ and $i^-$, are just inclusions of points so \eqref{eq:comp} tells us that the twisted cohomology is given by $x \in \mathbb{Z} $ with the relation $2 x=0$, and thus is isomorphic to $\mathbb{Z}_2$. It is clear that this result is the same for each choice of non-trivial $\theta$. As an immediate extension we get the equivalent result for any closed orientable surface, $\Sigma_g$, of genus $g$. Free homotopy classes of maps, ${\bf n}:\Sigma_g\to\mathbb{R} \mathbb{P}^2$, are given by an element $w_1(\mathbf{n}) \in H^1(\Sigma_g ; \mathbb{Z}_2) = \mathbb{Z}_2^{2g}$; if this is trivial they are distinguished by an unsigned degree ($\mathbb{N}$), while if it is non-trivial there are only two distinct classes ($\mathbb{Z}_2$). 

Figure \ref{fig:torus_homotopy} shows a diagrammatic version of this computation. Elements in the twisted homology $H_0(\Omega; \mathbb{Z}^{w_1})$ can be described by the total charge of a set of point charges on the torus whose sign flips upon passing through $\mathcal{B}$. As shown in the figure, this establishes an equivalence between $q_1+q_2$ and $q_1-q_2$, for any $q_1,q_2 \in \mathbb{Z}$, and so gives the topological classification as $\mathbb{Z}_2$.

\subsection{Computation for an arbitrary set of defects}

Now we will compute the homotopy classes of nematic textures on the complement of an arbitrary defect set. We first consider a nematic texture in $\mathbb{R}^3$ with a prescribed set of line defects, $L$, and no other singularities, so that $\mathbf{n}$ is nonorientable around each component of $L$. Furthermore, we will assume that $\textrm{lim}_{|x| \to \infty} \mathbf{n}(x) = \mathbf{n}_0$, a constant -- we will relax this assumption later. Our assumption that $\mathbf{n}$ is uniform at infinity allows us to compactify the domain and consider the domain $\Omega=S^3-N(L)$. A branch set, $\mathcal{B}$, for this domain is a Seifert surface for $L$, as shown in Figure \ref{fig:branch}. The double cover $\widehat{\Omega}$ is now the double cyclic cover~\cite{Lickorish}. Since $\mathcal{B}$ is a surface with boundary, $H_0(\mathcal{B}, \partial \mathcal{B})=0$ and so we use \eqref{eq:comp}. We are thus required to compute $H_{1}( \Omega - \mathcal{B}, \partial \Omega) / R$. 

\begin{figure}[t]
\begin{center}
\includegraphics{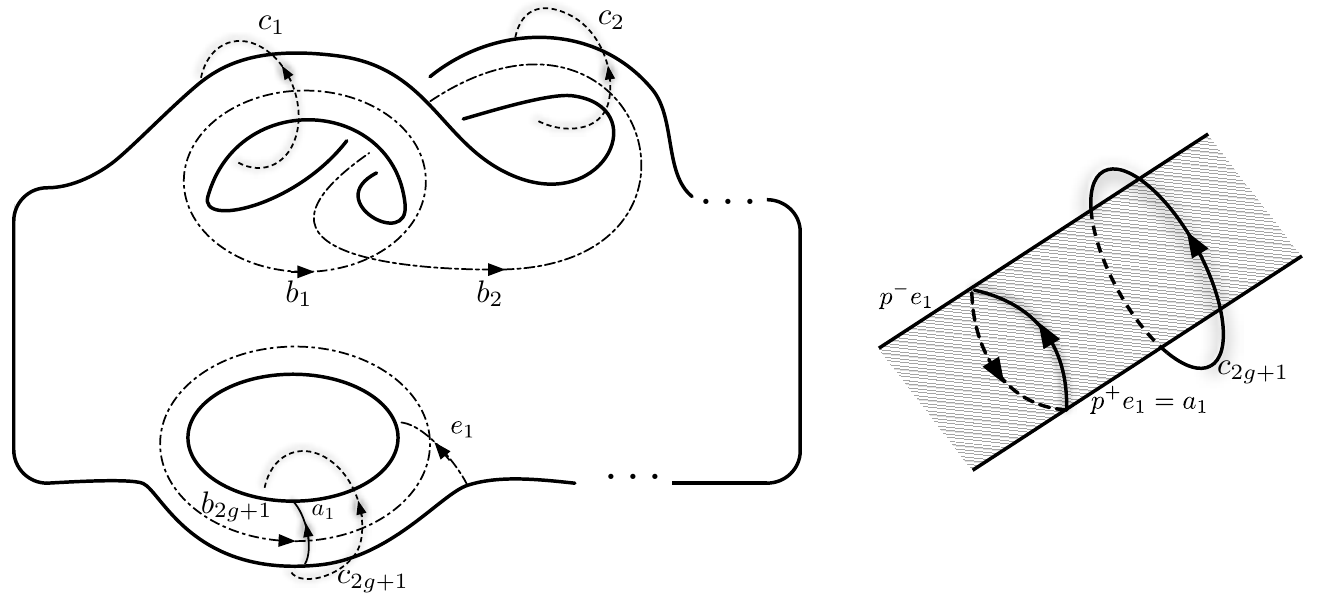}
\caption[Various cycles relating to $\mathcal{B}$ and $L$.]{\textit{Left}: Various cycles relating to $\mathcal{B}$ and $L$. $\mathcal{B}$ is represented as a surface with $g$ sets of double handles along the top and $|L|-1$ sets of loops along the bottom. The basis $a$ gives set of tethers that connect the link components. The cycles $b$ gives a basis for $H_1(\mathcal{B})$, and the cycles $\{b_i \} \cup \{e_j \}$ for $i \in [1, 2g]$ and $j \in [1, |L|-1]$ give a basis for $H_1(\mathcal{B}, \partial \mathcal{B})$. The basis $c$ generates $H_1(S^3-L)$. $\{a_i\} \cup \{c_i\}$ gives an overcomplete basis for $H_1(S^3-\mathcal{B}, \partial \mathcal{B})$. The bases $\{b_i\}$ and $\{c_i\}$ are chosen such that $\textrm{Lk}(c_i,b_j)=\delta_{ij}$. \textit{Right}: The relationship $2 a_i = c_{2g+i}$.}
\label{fig:ch4surfacebasis2}
\end{center}
\end{figure}

To do so, we make use of the cycles shown in Figure \ref{fig:ch4surfacebasis2}. The cycles $\{b_{i}\}$, $i=1,\dots,2g+|L|-1$, give a basis for the homology of the branch set $H_1({\cal B};\mathbb{Z})$, while the cycles $\{b_i \} \cup \{e_j \}$ for $i=1,\dots,2g$ and $j=1,\dots,|L|-1$ give a basis for the relative homology $H_1(\mathcal{B}, \partial \mathcal{B};\mathbb{Z})$. The cycles $\{c_i\}$, for $i=1,\dots,2g+|L|-1$, give a basis for the first homology of the complement $H_1(\Omega-{\cal B};\mathbb{Z})$, dual to the $\{b_i\}$. Finally, the relative cycles $\{a_i\}$, for $i=1,\dots,|L|-1$, may be thought of as positive push-offs of the $\{e_i\}$, $a_i=p^{+}e_i$, which together with the $\{c_j\}$ give an overcomplete basis for the relative homology of the complement $H_1(\Omega-{\cal B},\partial\Omega;\mathbb{Z})$. The redundancy in the description is removed by the set of relations, denoted $B$, obtained by setting cycles corresponding to each of the longitudes of the link components to zero. These can be written in terms of cycles on the surface as $p^+ b_i =0$ (equivalently $p^- b_i=0$), where $2g+1 \leq i \leq 2g+|L|-1$. 

We obtain the additional relations $R$ in \eqref{eq:comp} by taking into account the push-off of cycles in $H_1(\mathcal{B}, \partial \mathcal{B})$. $H_1(\mathcal{B}, \partial \mathcal{B}) = \mathbb{Z}^{2g} \oplus \mathbb{Z}^{|L|-1}$, where the first factor is generated by the first $2g$ $b_i$ in Figure \ref{fig:ch4surfacebasis2} and the second factor is generated by the $e_i$. Correspondingly there are two sets of relations, denoted $A$ and $C$. The relations of type $A$  come from setting $p^+ b_{i} + p^- b_{i}= 0$. Relations of type $C$ are written as $p^+ e_{i} + p^- e_{i}= 0$. From Figure \ref{fig:ch4surfacebasis2} we conclude that these relations take the form $2 a_i = c_{2g+i}$. 

We have $2g$ relations of type $A$, and $|L|-1$ relations of type $B$ and $C$, which can be written as
\begin{equation}
A_i := b^+_i+b^-_i = \sum_{j=1}^{2g+|L|-1} \textrm{Lk}(b_j,b_i^+)+\textrm{Lk}(b_j,b_i^-)=0
\end{equation}
\begin{equation}
B_i: = b^+_{2g+i} = \sum_{j=1}^{2g+|L|-1} \textrm{Lk}(b_j,b_{2g+i}^+)=0
\end{equation}
\begin{equation}
C_i := 2 a_i - c_{2g+i}=0
\end{equation}
and combined into a block matrix presenting $H^2(\Omega ; \mathbb{Z}^{w_1})$, where the horizontal blocks correspond to the first $2g$ $\{c_i\}$, the last $|L|-1$ $\{c_i \}$ and the $\{a_i\}$ and the vertical blocks correspond to relationships of type $A$, $B$ and $C$, as
\begin{equation}
\begin{pmatrix}
A^{(1)} & A^{(2)} & 0 \\
B^{(1)} & B^{(2)} & 0 \\
0 & 1 & -2
 \end{pmatrix}.
\end{equation}
By taking into account relationships of type $C$, it is clear that an equivalent group is presented by the matrix
\begin{equation}
M_1=\begin{pmatrix}
A^{(1)}  & 2A^{(2)} \\
B^{(1)} & 2B^{(2)} \\
\end{pmatrix}.
\label{eq:ch4pres1}
\end{equation}
Now let $S$ be the Seifert matrix for the link $L$, then in our notation the matrix
\begin{equation}
M_2=S+S^T=\begin{pmatrix}
A^{(1)} & A^{(2)} \\
2 B^{(1)} & 2 B^{(2)}
\end{pmatrix}
\label{eq:ch4pres2}
\end{equation}
is a presentation matrix for the first homology of the branched double cover of $S^3-L$ \cite{Lickorish}, which we denote $H_1(\Sigma(L))$. $M_1$ and $M_2$ are similar matrices with identical diagonal elements and it follows that they present the same group. From this we observe that the number of topologically distinct nematic textures associated to the defect set $L$ is given by elements of $H_1(\Sigma(L)) / x\sim -x$, where the equivalence relation comes, as usual, from the lack of orientation of the director. If $L$ is a split link, then it is a standard result~\cite{Lickorish,Rolfsen} that for a link $L$ with $N$ split components $L_i$ 
\begin{equation}
H_1(\Sigma(L))=\mathbb{Z}^{N-1} \oplus \bigoplus_i H_1(\Sigma(L_i)).
\end{equation}
The factors of $\mathbb{Z}$ at the front can be thought of as corresponding to the point-defect charges of the split components. There are $N-1$ of them rather than $N$ because our uniform boundary conditions demand that the total charge is zero, which reduces the degrees of freedom by one. We can lift this condition by supposing that there is a point defect `at infinity' in $S^3$, the charge of this point defect compensates for the charges of all the other defects. Finally we can also add an arbitrary number of point defects into our system. Doing this we obtain the following result. Let $\mathbf{n}$ be a director field for a nematic liquid crystal in $\mathbb{R}^3$ with a defect set $\mathcal{D}=\mathcal{P} \cup \mathcal{L}$, where $\mathcal{P}$ is the set of point defects and $\mathcal{L} = \cup_i L_i$ is the set of line defects, with each $L_i$ a non-split link or knot. Then the topology of the texture $\mathbf{n}$ is given by an element of the set
\begin{equation}
\left(\bigoplus_{p_i \in \mathcal{P}} \mathbb{Z} \right) \oplus \left ( \bigoplus_{L_j \in \mathcal{L}} \big (\mathbb{Z} \oplus H_1(\Sigma(L_j))  \big) \right) \Big/ x \sim -x.
\label{eq:ch4goal}
\end{equation}

The group $H_1(\Sigma(L))$, and consequently the homotopy classes of textures on the complement of knots and links, holds a considerable amount of richness. It can be computed in a variety of ways~\cite{Rolfsen, Lickorish}, with perhaps the simplest being through the Goeritz matrix associated to any regular projection. For a knot $K$ the order of the group $|H_1(\Sigma(K))|$ is always a finite odd integer, known as the determinant of the knot. For a link there is a richer set of phenomena, where $|H_1(\Sigma(L))|$ is either even or infinite. This is illustrated in table \ref{tab:ch4tab1} which shows $H_1(\Sigma(L))$ for the $(p,q)$ torus links with $p,q\leq 12$ (which is a knot if $\textrm{gcd}(p,q)=1$), counting pointed homotopy classes of nematic textures on the complement of the link. There are three properties of these groups one should observe: there are complex knots for which $|H_1(\Sigma(K))|=1$; there are some links for which $|H_1(\Sigma(L))|=\infty$ and in the case that $|H_1(\Sigma(L))|$ is finite, it is even rather than odd as in the case of knots. Links for which $|H_1(\Sigma(L))|=\infty$ have the interesting geometric property of admitting disconnected spanning surfaces~\cite{Lickorish}. This is trivially true in the case of a split link, but less obvious in general~\cite{note8}. To see this note that if a link supports disconnected spanning surfaces, then these may serve as a branch set for a nematic texture. As such, $\mathbf{n}$ will be orientable when restricted to the boundary of a thickened copy of one spanning surface component, $\mathcal{B}_a$. $\mathbf{n} \big |_{\mathcal{B}_a}$ will then be orientable, and one may compute a degree for this map (up to a sign). These degrees correspond to the integer summands. The last property, that for links $|H_1(\Sigma(L))|$ may be even, is the most physically relevant for nematic textures. As we have discussed previously~\cite{machon14}, the distinct homotopy classes of nematic textures on the complement of knots correspond to entangling the knot with a `Skyrmion tube' in a non-trivial way, or equivalently creating a tether with a double twist cylinder cross-section that connects two parts of the knot. For links, not all the textures have this interpretation. Links with elements of order 2 in $H_1(\Sigma(L))$ possess multiple topologically distinct planar textures. 
\begin{table}
\scriptsize{
\begin{center}
\begin{tabular}{c| c c c c c c c c c c c c}
\hline \hline \\ $p \setminus q$ & 2 & 3 & 4 & 5 & 6 & 7 & 8 & 9 & 10 & 11 & 12 \\ \hline 
2 & 2 & 3 & 4 & 5 & 6 & 7 & 8 & 9 & 10 & 11 & 12 \\
3 & 3 & $2^2$ & 3 & 1 & $ \mathbb{Z}^2$ & 1 & 3 & $2^2$ & 3 & 1 & $ \mathbb{Z}^2$ \\
4 & 4 & 3 & $2 \times  \mathbb{Z}^2$ & 5 & 12 & 7 & $ 4 \times  \mathbb{Z}^2$ & 9 & 20 & 11 & $6 \times \mathbb{Z}^2$ \\
5 & 5 & 1 & 5 & $2^4$ & 5 & 1 & 5 & 1 & $ \mathbb{Z}^4$ & 1 & 5 \\
6 & 6 & $ \mathbb{Z}^2$ & 12 & 5 & $2 \times \mathbb{Z}^4$ & 7 & 24 & $3 \times \mathbb{Z}^2$ & 30 & 11 & $4 \times \mathbb{Z}^4$ \\
7 & 7 & 1 & 7 & 1 & 7 & $2^6$ & 7 & 1 & 7 & 1 & 7\\
8 & 8 & 3 & $4 \times \mathbb{Z}^2$ & 5 & 24 & 7 & $2 \times \mathbb{Z}^6$ & 9 & 40 & 11 & $12 \times \mathbb{Z}^2$\\
9 & 9 & $2^2$ & 9 & 1 & $3 \times \mathbb{Z}^2$ & 1 & 9 & $2^8$ & 9 & 1 & $3 \times \mathbb{Z}^2$ \\
10 & 10 & 3 & 20 & $ \mathbb{Z}^4$ & 30 & 7 & 40 & 9 & $2 \times \mathbb{Z}^8$ & 11 & 60 \\
11 & 11 & 1 & 11 & 1 & 11 & 1 & 11 & 1 & 11 & $2^{10}$ & 11 \\
12 & 12 & $ \mathbb{Z}^2$ & $6 \times \mathbb{Z}^2$ & 5 & $4 \times \mathbb{Z}^4$ & 7 & $12 \times \mathbb{Z}^2$ & $3 \times \mathbb{Z}^2$ & 60 & 11 & $2 \times \mathbb{Z}^{10}$ \\
\hline \hline
\end{tabular}
\end{center}
}
\caption[$H_1(\Sigma(L))$ for $(p,q)$ torus links with $2 \leq (p,q) \leq 12$.]{$H_1(\Sigma(L))$ for $(p,q)$ torus links with $2 \leq (p,q) \leq 12$. $x^n$ implies a group $(\mathbb{Z}_x)^n$, integer summands are given as usual. When $\textrm{gcd}(p,q)=1$, one obtains a knot and $|H_1(\Sigma(L))|$ is given by the Alexander polynomial $(t^{pq}-1)(t-1)/(t^p-1)(t^q-1)$ evaluated at -1.}
\label{tab:ch4tab1}
\end{table}

\section{Planar Textures}
\label{sec:planar}

Nematic textures on the complement of the Hopf link have the property that they may all be brought into a planar form, with the director lying everywhere in the $xy$-plane. Explicitly, there are two distinct homotopy classes, representatives for which may be given in the form 
\begin{equation}
{\bf n} = \bigl( \cos(\phi/2), \sin(\phi/2), 0 \bigr) ,
\label{eq:planar_n}
\end{equation} 
where $\phi$ is the argument of a simple polynomial function, for instance 
\begin{align}
\phi & = \textrm{Arg}\, \bigl( (x+i)^2 +(y-1)^2+ (z-i)^2 \bigr)\bigl( (x-i)^2 +(y+1)^2 +(z-i)^2 \bigr) ,
\label{eq:Hopf_1}
\intertext{and} 
\phi & = \textrm{Arg}\, \bigl( (x+i)^2 +(y-1)^2+ (z-i)^2 \bigr)\bigl( (x+i)^2 +(y+1)^2 +(z+i)^2 \bigr) ,
\label{eq:Hopf_2}
\end{align}
which come from the Milnor fibration of the Hopf link complement~\cite{Milnor,machon14,MachonThesis}. The two functions differ only by conjugation of the second factor. In this example the two components of the Hopf link correspond to the curves $(\pm\cos(t),\pm 1\pm\sqrt{2}\sin(t),\cos(t))$ and one may check that the surface $\phi=\pi$ induces linking number $+1$ between them in the first case, \eqref{eq:Hopf_1}, and linking number $-1$ in the second, \eqref{eq:Hopf_2}. 

The arguments of each of the two factors in \eqref{eq:Hopf_1}, $\theta_1$ and $\theta_2$, say, define angles that wind around each of the disclination lines and exhibit the homotopy equivalence of the Hopf link complement with the torus, $T^2$, mentioned in the introduction~\cite{note9}. Thus, the two Hopf link textures are equivalent to the maps $T^2\to\mathbb{RP}^2$ of the form~\eqref{eq:planar_n} with $\phi=\theta_1\pm\theta_2$. In these textures the director is non-orientable around both the meridian and longitude of the torus. When it is only non-orientable around one of them, as is the case for a Janus colloid~\cite{cavallaro13,machon13}, there are still only two homotopy classes of maps but again each of these has a planar representative; examples for each class are the textures of the form~\eqref{eq:planar_n} with $\phi=\theta_1$ and $\phi=\theta_1+2\theta_2$. Indeed, a little work shows that all non-orientable textures on the torus may be brought into such a planar form~\cite{janich87}. By contrast, for textures that are orientable, which are classified by an unsigned degree, the only charge $q \in \mathbb{N}$ for which the texture may be made planar is the trivial case, $q=0$. All others are essentially non-planar. It may be noted that the surface normal to a toroidal colloidal particle, corresponding to homeotropic anchoring, has degree 0 and so is in the planar class. 

This dichotomy between planar and non-planar textures represents a fundamental feature of the global theory of nematic defects. As the neighbourhood of any disclination loop is a torus, and the texture on such a torus can be made planar~\cite{janich87}, the local texture in a neighbourhood of the disclinations is not sensitive to the dichotomy of the global theory. That this is a genuine topological feature of the global theory of nematics follows from the topological characterisation we provide here: namely, the number of planar textures supported by a given link is equal to the number of elements in the order 2 subgroup of $H_1(\Sigma(L))$. Note that this gives a unique planar texture on the complement of any knotted defect line. Some links also have unique planar textures; an example in table \ref{tab:ch4tab1} is the $(5,10)$ torus link, which has an infinite number of homotopy classes of textures but only one of them admits a planar representative. Conversely, some links have planar representatives for all of their homotopy classes; examples from table \ref{tab:ch4tab1} are the $(n,n)$ torus links for $n$ odd, which have $2^{n-1}$ distinct planar textures and none that are not planar. 
We now proceed to establish this result. 

Any planar texture is a map of the form $\Omega\to\mathbb{RP}^1$. Of course, such a texture can also be viewed as a map into $\mathbb{RP}^2$ since $\mathbb{RP}^1$ is a subset of $\mathbb{RP}^2$. The question, then, is which homotopy classes of maps $\Omega\to\mathbb{RP}^2$ are obtained in this way? The question may be turned around and phrased equivalently as asking for the nature of any obstruction to compressing a map $\Omega\to\mathbb{RP}^2$ to a planar one, $\Omega\to\mathbb{RP}^1$. This question is addressed directly. Given any nematic texture ${\bf n}:\Omega\to\mathbb{RP}^2$ we try to compress it to a planar one in an inductive manner proceeding dimension-by-dimension on a cell-decomposition for the domain. It is not hard to see that every texture compresses over the 1-skeleton, so the first obstruction arises with the behaviour on the 2-cells~\cite{note10}. Having homotoped the director so that it is planar on the 1-skeleton, the texture on each 2-cell corresponds to a map from a disc into $\mathbb{RP}^2$ such that its boundary lies in an equatorial $\mathbb{RP}^1$, {\it i.e.} $(D^2,\partial D^2)\to(\mathbb{RP}^2,\mathbb{RP}^1)$, and hence to an element of the relative homotopy group $\pi_2(\mathbb{RP}^2,\mathbb{RP}^1)$. If the texture is actually planar (and not simply homotopic to a planar one) then this element is trivial. Thus every map $\Omega\to\mathbb{RP}^2$ gives rise to a 2-cocycle with local coefficients in $\pi_2(\mathbb{RP}^2,\mathbb{RP}^1)$. Under homotopy this 2-cocycle will change but its cohomology class does not. For any planar texture this class is trivial and the obstruction to compressing any texture to a planar one is characterised by a degree 2 cohomology class with local coefficients in the relative homotopy group $\pi_2(\mathbb{RP}^2,\mathbb{RP}^1)$. 

\begin{figure}
\begin{center}
\includegraphics{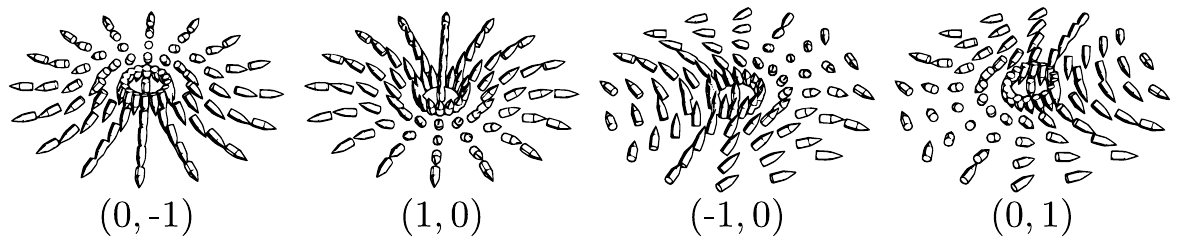}
\caption{Non-trivial elements of $\pi_2(S^2,S^1) = \mathbb{Z}^2$ and charge fractionalisation of Skyrmions. Elements are specified by two integers, $(p,q)$ which give the number of times the map wraps around the northern and southern hemispheres. The winding in the equatorial $S^1$ around the boundary is given by $p-q$. The antipodal map sends $(p,q)$ to $(-q,-p)$ which preserves the winding on the boundary as $x\to -x$ is orientation preserving in two dimensions.}
\end{center}
\label{fig:rel}
\end{figure}

To describe the local coefficient system note that any map $D^2 \to \mathbb{R} \mathbb{P}^2$ must be orientable which gives the isomorphism (as groups) $\pi_2(\mathbb{R} \mathbb{P}^2, \mathbb{R} \mathbb{P}^1) = \pi_2(S^2, S^1) = \mathbb{Z}^2$~\cite{note11}. These two integers can be thought of as counting the number of times the map wraps around the northern and southern hemispheres respectively, with the winding number in $S^1$ around the boundary given by their difference. This can be viewed as a type of charge fractionalisation, illustrated in Figure \ref{fig:rel}, where a whole Skyrmion is split into two halves (merons), corresponding to coverings of the two hemispheres. In the liquid crystalline case one must also take into account the action of $\pi_1(\mathbb{R} \mathbb{P}^2)$ on $\pi_2(\mathbb{R} \mathbb{P}^2, \mathbb{R} \mathbb{P}^1)$ which sends a positively oriented northern hemisphere to a negatively oriented southern hemisphere and so on, and so sends $(a,b)$ to $(-b,-a)$. This gives $\pi_2(\mathbb{R} \mathbb{P}^2, \mathbb{R} \mathbb{P}^1)$ the structure of the group ring $\mathbb{Z}[\mathbb{Z}_2]$. 

The homotopy of any nematic texture ${\bf n}:\Omega\to\mathbb{RP}^2$ to be planar on the 1-skeleton gives a homomorphism of twisted cohomology groups 
\begin{equation}
\rho: H^2(\Omega ; \mathbb{Z}^{w_1}) \to H^2(\Omega ; \mathbb{Z}[\mathbb{Z}_2]^{w_1}),
\end{equation}
in which the planar textures correspond to the kernel. We thus need to compute the group $H^2(\Omega ; \mathbb{Z}[\mathbb{Z}_2]^{w_1})$ and the map $\rho$. $H^2(\Omega ; \mathbb{Z}[\mathbb{Z}_2]^{w_1})$ admits the same decomposition through the covering space as $H^2(\Omega; \mathbb{Z}^{w_1})$. On the branch set, $\mathcal{B}^{w_1}$, and the sheets of the cover, $\Omega_1$ and $\Omega_2$, the coefficient system is trivial and given by $\mathbb{Z}^2$. The equivariance condition now requires elements in \eqref{eq:mayer} to be of the form $((x_1,x_2),(-x_2,-x_1)) \in H_{1}(\mathcal{B}^{w_1}, \partial \mathcal{B}^{w_1}; \mathbb{Z}^2)$ and so on. It follows that 
\begin{equation}
H^2(\Omega ; \mathbb{Z}[\mathbb{Z}_2]^{w_1}) =  \left ( \bigoplus_{j=1}^2 H_{1}( \Omega - \mathcal{B}, \partial \Omega) \right ) / P.
\end{equation}
where the relations $P$ are of the form
\begin{equation}
\begin{pmatrix} i^+ & i^- \\ i^- & i^+
\end{pmatrix} \begin{pmatrix} x_1 \\ x_2 \end{pmatrix} = 0,
\label{eq:newrel}
\end{equation}
and the structure of the maps $i^+$ and $i^-$ is inherited from \eqref{eq:mayer}. 

\begin{figure}
\begin{center}
\includegraphics{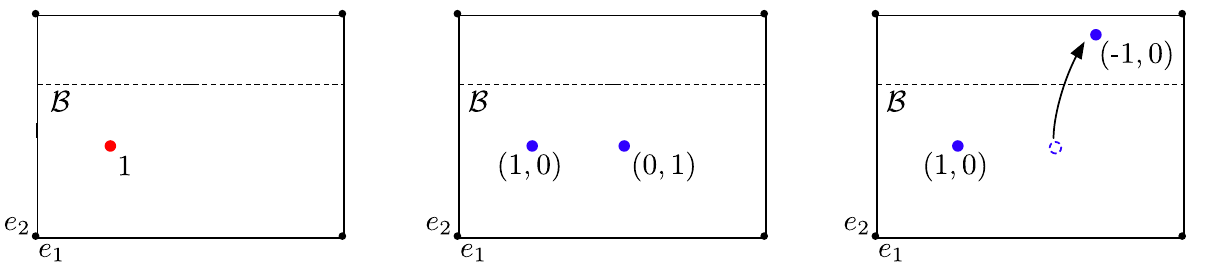}
\caption{Diagrammatic calculation of planar textures on the torus. A Skyrmion of charge $1$ is split via the relative homotopy group into $(1,1)=(1,0)+(0,1)$. Moving the $(0,1)$ half-Skyrmion through $\mathcal{B}$ sends $(0,1) \to (-1,0)$, which can be annihilated with the remaining $(1,0)$, resulting in a planar texture.}
\label{fig:torus_homotopy_2}
\end{center}
\end{figure}

To define the map $\rho$, we first select any planar texture ${\bf n}_0$ to represent the element $0$ in $H^2(\Omega;\mathbb{Z}^{w_1})$. Homotopically distinct modifications of $\mathbf{n}_0$, that do not alter the behaviour on the 1-skeleton, are achieved via the addition of Skyrmions to 2-cells. The addition of a degree one Skyrmion to a given 2-cell corresponds to the addition of a cooriented northern and southern hemisphere (the element $(1,1)$ in $\pi_2(\mathbb{RP}^2, \mathbb{RP}^1)$). It follows that $\rho$ acts as the diagonal map, sending $x \to (x,x)$ and that the texture is planar if $(x,x) \sim (0,0)$ under the relations \eqref{eq:newrel}, which can be written as
\begin{equation}
\begin{pmatrix}
x \\ x
\end{pmatrix} \sim  \begin{pmatrix} x \\ x \end{pmatrix} + \begin{pmatrix} i^+ & i^- \\ i^- & i^+
\end{pmatrix} \begin{pmatrix} \alpha \\ \beta \end{pmatrix},
\end{equation}
for any $\alpha,\beta\in H_1({\cal B},\partial{\cal B};\mathbb{Z})$. Setting $\beta=0$ we find $(x,x) \sim (x+i^+ \alpha,x+i^- \alpha)$. It is always possible to find an $\alpha$ such that $i^- \alpha = -x$ so long as $x$ represents a torsion class in $H^2(\Omega ; \mathbb{Z}^{w_1})$, giving $(x,x) \sim (x+i^+ \alpha, 0)$. Hence the map is planar if $x+i^+ \alpha = 0$ or 
\begin{equation}
(i^++i^-)\alpha=-2x,
\end{equation}
which is equivalent to the statement that $2x=0 \in H^2(\Omega ; \mathbb{Z}^{w_1})$. From this we obtain the result that $\mathbf{n}$ is homotopic to a planar texture if its class $[\mathbf{n}] \in H^2(\Omega ; \mathbb{Z}^{w_1})$ is an element of order 2. It is clear that this does not depend on which planar texture ${\bf n}_0$ is used as the reference map that represents the 0 element, and so one finds that planar textures are in correspondence with the order 2 subgroup of $H^2(\Omega ; \mathbb{Z}^{w_1})$. Note that because the order 2 elements are those which are invariant under the transformation $x \to -x$, there is no additional reduction when passing to free homotopy classes of planar textures. As in the homotopy classification, one can give a diagrammatic version of this construction, illustrated in Figure~\ref{fig:torus_homotopy_2}. The situation is identical to Figure \ref{fig:torus_homotopy}, but now the point charges are split (fractionalised) through the relative homotopy group, into northern hemispheres and southern hemispheres. 

For links the physical interpretation of these planar textures cannot be given in terms of Skyrmion-like distortions entangled with the link for, being planar, they have none. A planar texture on the complement of a link $L$ may be written as $\mathbf{n} = (\cos \phi, \sin\phi, 0)$, for $\phi: \Omega \to \mathbb{R} \mathbb{P}^1$. The preimage of a particular orientation gives an orientable spanning surface for the link. Distinct planar textures are therefore associated with spanning surfaces for a particular link that are not cobordant as $X$-submanifolds~\cite{note12}. In the case of the Hopf link, these have a particularly simple interpretation: an orientable surface spanning a link induces a well defined relative orientation on the link components, which allows one to compute an unambiguous linking number. The two planar textures on the Hopf link can therefore be interpreted as inducing different linking numbers through computation of a preimage surface. In general, however, we do not know of the full distinction between planar textures~\cite{note13}. 

We remark in closing that the fractionalisation of Skyrmions that we described here occurs also in magnetic systems~\cite{lin15} and is analogous to the nature of vortex cores in $^3$He-A~\cite{mermin78}. The group $\pi_2(S^2,S^1)$ classifies nonsingular topological objects in a system whose order parameter lives on $S^2$, but for which there is an equatorial $S^1$ with lower energy. For example, take the classical energy functional for the unit magnetisation, $\mathbf{m}$, of a two-dimensional ferromagnet with easy-plane anisotropy
\begin{equation}
\int |\nabla \mathbf{m}|^2 + A(\mathbf{m} \cdot \mathbf{e}_z)^2 d^2 x.
\end{equation}
If $A$ is large then the groundstate manifold is the equatorial $S^1 \subset S^2$ with zero $z$-component. Defects in this system are then described by a winding number $q \in \pi_1(S^1)$. The singular core of such an $S^1$ defect can be continuously filled by allowing $\mathbf{m}$ to vary over the full sphere (`escape in the third dimension'), and thus defines an element of the relative homotopy group $\pi_2(S^2,S^1)$. The situation we describe here generalises this by having textures on a general three-dimensional domain, rather than a thin film, and the non-trivial action of $\pi_1$ on $\pi_2$ that occurs for nematic order.

\section{Physical Realisation of Homotopy Classes}

Several techniques have been developed for determining the director field in complicated three-dimensional textures, such as three-photon excitation fluorescence polarising microscopy, confocal microscopy, or coherent anti-Stokes Raman scattering microscopy. The topological properties of the director field can then be deduced using the Pontryagin-Thom construction for nematic textures~\cite{chen13,BryanThesis}. We describe here how this topological characterisation applies to recent experiments of Tasinkevych, Campbell and Smalyukh on toroidal nematic drops~\cite{tasinkevych14} and how it relates to the global defect topology we have presented. These experiments create a variety of fascinating nematic textures with linked and knotted disclination lines in polymer stabilised nematic drops with the topology of a handlebody (genus between 1 and 5). We focus on the example of the Hopf link created in a solid torus (genus 1). 

The Pontryagin-Thom construction represents the topology of the texture by a coloured surface~\cite{BryanThesis}. It states that there is a bijection between homotopy classes of maps $\Omega \to \mathbb{R} \mathbb{P}^2 = X \cup \{pt \}$, where $X$ is a line bundle over $\mathbb{RP}^1$, and cobordism classes of $X$-submanifolds of $\Omega$, where a $X$-submanifold is a codimension one submanifold with a bundle map from its normal bundle into $X$. In practical terms, one plots the preimage of an equatorial $\mathbb{R} \mathbb{P}^1 \subset \mathbb{R} \mathbb{P}^2$, which gives a surface, that we call a PT surface. The additional angular degree of freedom in the $\mathbb{R} \mathbb{P}^1$ can be visualised by colouring the PT surface. The resulting coloured surface captures the topological information in the texture. Homotopies of the texture are equivalent to cobordisms between surfaces that are consistent with the colouring and bundle map. 

\begin{figure}
\begin{center}
\includegraphics{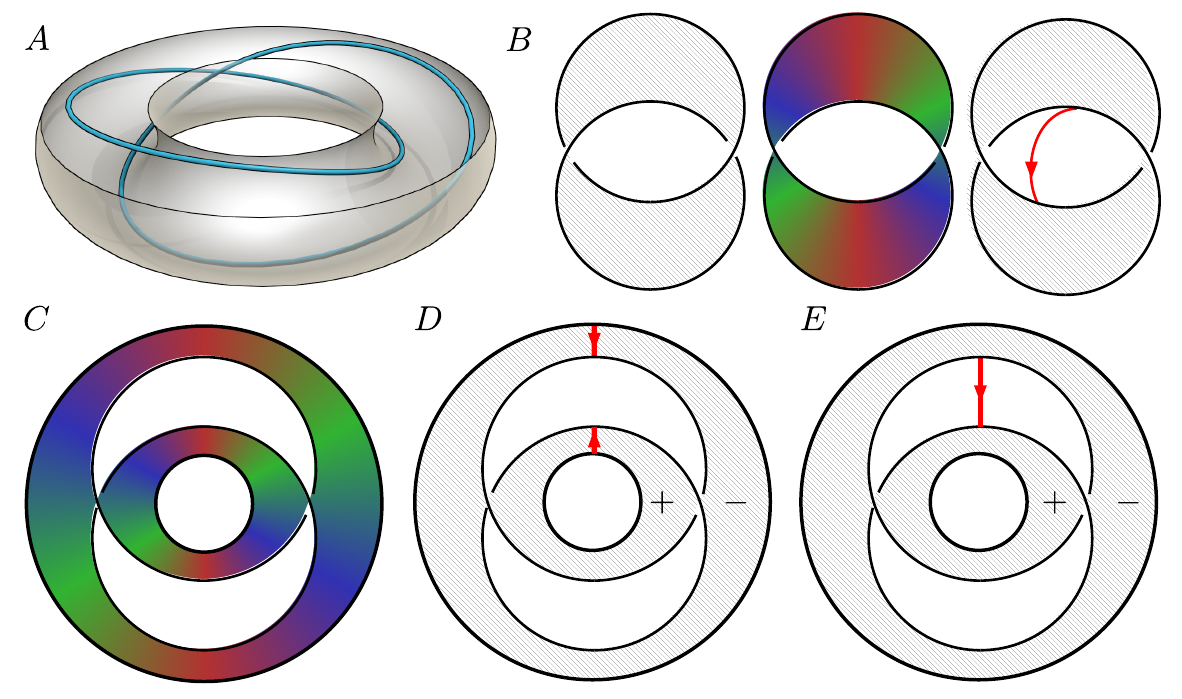}
\caption{(A) Cutaway cartoon of the Hopf link defect inside a toroidal droplet, as created in~\cite{tasinkevych14}. (B) PT representations of texture on the Hopf link. \textit{Left}: a planar texture with $\textrm{Lk}=+1$, corresponding to a solid colour surface. \textit{Middle}: a homotopic texture with a $\textrm{Lk}=-1$ surface with colour winding. \textit{Right}: Another homotopic texture, where the colour winding has been removed to form a tether connecting the link components. (C) Schematic PT surface inferred from experimental data~\cite{tasinkevych14}. (D) Compressing the winding into tethers and pushing them off the surface. The differing orientations arise due to the tethers being pushed off different sides of the surface, indicated by the $\pm$ signs. (E) Connecting the tethers by passing over the roof of the solid torus to form a single tether.}
\label{fig:big_fig}
\end{center}
\end{figure}

Using the PT construction allows one to understand how distinct homotopy classes of textures may be realised. Taking again the Hopf link, if the texture is planar then the PT surface will be of constant colour, as shown in Figure \ref{fig:big_fig}(B), and one may compute a linking number. The surface, however, can have colour winding. Suppose one had a surface inducing the opposite linking number, with a colour winding of, say, $2 \pi$, as shown in the Figure, then this colour winding may be compressed until it is located in just a small region of the surface, with the rest of the surface of constant colour. The colour winding can then be `pulled off' to form an additional surface which has the form of a tether (or Skyrmion tube) connecting the link components, oriented by the direction of the colour winding along its meridian. This defines a relative homology cycle and consequently an element of $H_1(\Omega - \mathcal{B}, \partial \Omega)$, and so an element of $H^2(\Omega ; \mathbb{Z}^{w_1})$. Addition of a $ 2\pi$ colour winding or, equivalently, a tether corresponds to an alteration of the homotopy class of the texture by the homology class of the tether in the group $H_1(\Omega, \partial \Omega ; \mathbb{Z}^{w_1})=H^2(\Omega ; \mathbb{Z}^{w_1})$. In the case of the Hopf link, this simply means that addition of a tether to a planar texture of one linking number is homotopic to a planar texture with the other linking number. This homotopy is constructed explicitly by Chen~\cite{BryanThesis} for the torus, which is homotopy equivalent to the complement of the Hopf link.

We are now in a position to show how the PT construction may be applied to understand experimental data. We will use an example of a Hopf link defect inside a toroidal droplet with homeotropic boundary conditions created experimentally by Tasinkevych, Campbell and Smalyukh~\cite{tasinkevych14}, shown in Figure \ref{fig:big_fig}(A). The first step is to compute the first invariant $w_1(\mathbf{n})$. In this case  $H^1(\Omega ; \mathbb{Z}_2) = \mathbb{Z}_2^3$, one factor for the meridian of each defect line and a third for the cycle that goes around the interior of the droplet. $\mathbf{n}$ is non-orientable around each of the defect lines, but not around the droplet, so $w_1(\mathbf{n}) = (1,1,0)$. Now we must compute the twisted cohomology group $H^2(\Omega;\mathbb{Z}^{w_1})$. A branch set for the system is given by a Seifert surface for the Hopf link defects, this is just an annulus so we get one relation. $H_1(\Omega - \mathcal{B}; \partial \Omega)=\mathbb{Z}^2$, consisting of a tether, $e_1$ connecting the droplet to one defect line, and a tether $e_2$ connecting the defect lines. The relation from $\mathcal{B}$ sets $2 e_2=0$ and so one finds that $H^2(\Omega ; \mathbb{Z}^{w_1})$ is given by $\mathbb{Z} \oplus \mathbb{Z}_2$ and so free homotopy classes are given by $\mathbb{N} \oplus \mathbb{Z}_2$. The droplet has homeotropic boundary conditions and it follows that the degree of $\mathbf{n}$ on the boundary is $0=\chi(T^2)$, and so $\mathbf{n}$ represents 0 in the factor of $\mathbb{N}$. All that remains is to determine the $\mathbb{Z}_2$ invariant. Figure \ref{fig:big_fig}(C) shows the PT surface for this system inferred from published results~\cite{tasinkevych14}. It contains a full $2\pi$ colour winding. Compressing this winding and pushing it off results in two small tethers, shown in Figure \ref{fig:big_fig}(D). Their orientations differ because they are pushed off different sides of the surfaces, indicated by $+$ and $-$ in the Figure. These tethers can be connected by passing over the roof of the droplet and combined, forming a single tether, shown in Figure \ref{fig:big_fig}(D). We thus find that, even though the surface has a linking number of $-1$, it has a tether, so is homotopic to a $\textrm{Lk}=+1$ planar texture.

\section{Nematic Order as a Vector Bundle}
\label{sec:vb}

A complementary perspective on nematic order comes from considering it as a vector bundle~\cite{machon16}. The $SO(3)$ symmetry of the isotropic phase is the structure group of the tangent bundle of the domain, $T \Omega$, with the low temperature group $D_\infty$ picking out a rank one subspace of $T \Omega$ that is preserved under the symmetry operations. Correspondingly one obtains a splitting of the tangent bundle into vectors along the nematic order and vectors orthogonal to the director as
\begin{equation}
T \Omega = L_\mathbf{n} \oplus \xi.
\label{eq:split}
\end{equation}
The low temperature group $D_{\infty}$ is the structure group of the bundle $\xi$, while the structure group of the line bundle $L_{{\bf n}}$ is $\mathbb{Z}_2$. One can then ask how the homotopy invariants of the nematic are reflected in the characteristic classes of these vector bundles. The first invariant $w_1(\mathbf{n})$ is simply the first Stiefel-Whitney class of both $L_\mathbf{n}$ and $\xi$, which describes the orientability of the bundles~\cite{note14}. Because $\xi$ is, in general, non-orientable it does not possess a Chern (or equivalently Euler) class. However, as in the case of the homotopy classes of nematic textures, it possesses a twisted Euler class. Given an oriented equivariant map on the covering space, $\hat{\mathbf{n}}: \Omega^{w_1} \to S^2$, the splitting \eqref{eq:split} also lifts as
\begin{equation}
T \Omega^{w_1} = L_\mathbf{\hat{n}} \oplus \hat{\xi},
\end{equation}
where now the bundles are orientable. Consider an equivariant vector field $\hat{\mathbf{m}}$, with $\hat{\mathbf{m}} \cdot \hat{\mathbf{n}} =0$. The zeros, $\hat{\mathbf{m}}^{-1}(0)$ form an equivariant set of codimension 2, and thus represent an equivariant homology cycle. They are topologically required to exist if this cycle is non-trivial or equivalently if the Poincar\'{e} dual $PD[ \hat{\mathbf{m}}^{-1}(0)] \in H^2(\Omega ; \mathbb{Z}^{w_1})$ is non-zero, which defines the twisted Euler class $e(\hat{\xi})$ of the bundle $\hat{\xi}$. It is a standard result~\cite{note15} that the twisted Euler class is given by 
\begin{equation}
e(\hat{\xi}) = 2 [\mathbf{n}] \in H^2(\Omega ; \mathbb{Z}^{w_1}).
\label{eq:eulertwice}
\end{equation}
Finally, to account for free homotopy classes, one should once again quotient by the equivalence relation $x \sim -x$. Thus one observes that the condition for the texture to be planar is equivalent to the statement that the twisted Euler class of the orthogonal bundle, $\xi$, also vanishes. This makes sense; if the texture can be made planar so that the director lies everywhere in the $xy$-plane then the vector that always points in the $z$-direction, ${\bf m}=(0,0,1)$, is a globally defined, nowhere zero section of the bundle $\xi$ and so its Euler class must vanish. Or, from another perspective, one can say that the planar textures are not captured by these characteristic classes. 

For systems with non-uniform ground states, such as cholesteric~\cite{beller14,machon16} and smectic~\cite{chen09} liquid crystals, information from the gradient tensor $\nabla \mathbf{n}$ is particularly important. In this case the twisted Euler class gives global constraints on the structure of the orthogonal gradient tensor, $\nabla_\perp \mathbf{n} = (\delta_{ij}-n_in_j) \nabla_j n_k$. In cholesterics zeros of the deviatoric part of $\nabla_\perp \mathbf{n}$ are features readily identifiable as the cores of $\lambda$ lines and double twist cylinders~\cite{beller14,machon16}, in geometric terms these are singularities in the pitch axis or umbilic lines. The twisted Euler class of the tensor bundle in which the deviatoric gradients live thus gives global constraints on these objects. In particular the Poincar\'{e} dual of the Euler class of the bundle gives a homology cycle that describes the topological class of all $\lambda$ lines in the system. Because $\nabla_\perp \mathbf{n}$ is a rank 2 tensor, the twisted Euler class of the associated tensor bundle is given by $4 [\mathbf{n}]$ following the standard rules of tensor products. It follows that the $\lambda$ lines in a cholesteric represent the homology cycle $PD \big [4 [\mathbf{n}] \big ]$. In the context of knotted and linked disclinations, this raises the interesting question of elements of order 4 in the group $H_1(\Sigma(L))$. Textures in this class do not admit a non-zero vector field orthogonal to $\mathbf{n}$ but do admit an orthogonal director field $\mathbf{m}$, or equivalently a non-zero traceless symmetric rank 2 tensor field $T_{ij}$, where $n_i T_{ij}=0$, whose positive eigenvector is $\mathbf{m}$. 

The pitch axis $\mathbf{p}$ in the traditional description of cholesterics, describing the principal local direction of twist in the order, is precisely such an orthogonal director field~\cite{kleman69,beller14}. As such, elements of order 4 are associated with cholesteric textures on the complement of a defect set for which the pitch axis is well defined everywhere, but is non-orientable. In terms of the traditional description of lines in cholesterics as $\chi$, $\lambda$, $\tau$ lines, this corresponds to a system containing $\chi$ lines (disclinations in $\mathbf{n}$) and $\tau$ lines (disclinations in $\mathbf{n}$ around which $\mathbf{p}$ is non-orientable) but no $\lambda$ lines (singularities in $\mathbf{p}$).

\begin{figure}
\begin{center}
\includegraphics{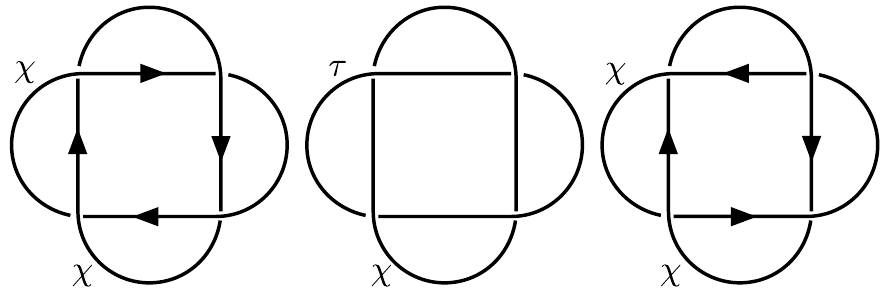}
\caption{Cholesteric textures on the complement of Solomon's seal. In this case $H_1(\Sigma(L))=\mathbb{Z}_4$. Under the equivalence relation $x \sim -x$, we have three distinct textures, $0$, $\{1,3\}$ and $2$. $0$ and $2$ are order two, and thus homotopic to planar textures which can be thought of as regular disclinations ($\chi$ lines) with differing relative orientations. The remaining texture is of order 4, and as such does not require the existence of $\lambda$ lines and can be represented by one $\chi$ line and one $\tau$ line.}
\label{fig:solomons}
\end{center}
\end{figure}

A simple example of such a situation, illustrated in Figure \ref{fig:solomons}, is the case of textures on the complement of the $(4,2)$ link, also known as Solomon's seal. In this case $H_1(\Sigma(L))=\mathbb{Z}_4 = \{0,1,2,3\}$. There are two elements of order two: $0$ and $2$, and two elements of order four: $1$ and $3$ which are identified under the equivalence relation $x \sim -x$. The elements of order two can be realised as planar textures and interpreted as corresponding to distinct relative orientations (and hence linking numbers) for the two disclinations. The element of order 4 corresponds to a system with one regular $\chi$ disclination line and one $\tau$ line. Note that exchange of the $\tau$ and $\chi$ labels can be accomplished by a homotopy which separates the $\tau$ line into a $\chi$ and $\lambda$ line, which is then merged with the other $\chi$ line to create a $\tau$ line.

\acknowledgements{We are grateful to Bryan Chen, Simon \v{C}opar and Randy Kamien for useful discussions. This work was supported in part by the UK EPSRC through Grants No. A.MACX.0002 (TM and GPA) and No. EP/N007883/1 (GPA). TM also partially supported by a University of Warwick Chancellor's International Scholarship and a University of Warwick IAS Early Career Fellowship.}


\begin{thebibliography}{99}
\bibitem{note1} There are 16 pointed homotopy classes but only 10 free classes after accounting for the equivalence $x\sim -x$. This does not affect the number of planar textures.
\bibitem{note2} Given a map $\theta: \pi_1(\Omega) \to \mathbb{Z}_2$, there is a unique map $w_1$, so that  $\theta = w_1 h$ in the sequence of maps $\begin{CD} \pi_1(\Omega) @> h >> H_1(\Omega) @> w_1>> \mathbb{Z}_2 \end{CD}$, where $h$, the Hurewicz map, is the Abelianisation of $\pi_1(\Omega)$.
\bibitem{note3} More generally, consider the sequence of maps $\begin{CD}\pi_1(\Omega^{w_1}) @> p^\ast>> \pi_1(\Omega) @ > \theta >> \mathbb{Z}_2\end{CD}$, where $p^\ast$ is the map on the fundamental group induced by the projection map, $p$, of the cover. Then the cover is defined so that $\textrm{Image}(p^\ast) = \textrm{Ker}(\theta)$.
\bibitem{note4} The equivariant cohomology $H^\bullet(\Omega^{w_1})_{eq}$ is the cohomology of the chain complex $C^\bullet(\Omega^{w_1})_{eq} \subset C^\bullet(\Omega^{w_1})$ consisting of cochains of the form $c- t^\sharp c$, where $t^\sharp$ is the map induced by the deck transformation of $\Omega^{w_1}$.
\bibitem{note5} Note that, in general, one does not have to choose $\mathcal{B}$ to be orientable, we only demand this for convenience. If we relax the assumption of orientability, then a class of branch sets can be obtained through the Pontryagin-Thom construction~\cite{chen13}. If $\alpha$ is an equatorial $\mathbb{R} \mathbb{P}^1 \subset \mathbb{R} \mathbb{P}^2$, then $\mathbf{n}^{-1}(\alpha)$ gives a codimension 1 set $\mathcal{B}_\alpha$ in $\Omega$. Any loop $\gamma$ with $\theta(\gamma)=1$ must intersect $\mathcal{B}_\alpha$ an odd number of times since $\alpha$ represents a non-trivial cycle in $\pi_1(\mathbb{R} \mathbb{P}^2)$, and any two non-trivial cycles on $\mathbb{R} \mathbb{P}^2$ generically intersect an odd number of times, so $\mathcal{B}_\alpha$ defines a branch set.
\bibitem{note6} If $\theta$ is trivial and $\mathbf{n}$ is orientable, $\mathcal{B}$ is the empty set. In this case $\Omega^{w_1}$ is two disjoint copies of $\Omega$, with $t$ swapping between sheets. Correspondingly the coefficient system $\mathbb{Z}^{w_1}$ is trivial, and we need only compute the regular cohomology of $\Omega$. In the case of $N$ point defects, $\Omega = \vee^N S^2$ and $H^2(\Omega; \mathbb{Z})=\mathbb{Z}^N$, with free homotopy classes then given by $\mathbb{Z}^N / x \sim -x$. This demonstrates the known fact that taking into account global information removes the ambiguity in adding point defect charges. One can fix the sign of only one point defect, with the charge of the others full integers in comparison.
\bibitem{note7} They come from applying Poincar\'e-Lefschetz duality to $H^{2}(\Omega;\mathbb{Z}^{w_1})$ and so are in co-degree 2.
\bibitem{note8} The authors realised this during an experiment with soap films. A soap film on the $(4,4)$ torus link, for which $|H_1(\Sigma(L))|=\infty$, will collapse until it forms two linked Hopf links.
\bibitem{note10} This is the only potential obstruction given that $\Omega$ is homotopy equivalent to a 2-complex.
\bibitem{note11} To compute $\pi_2(S^2,S^1)$, one takes the relevant short exact sequence in relative homotopy groups, $\begin{CD} 0 @>>> \mathbb{Z} @ >>> \pi_2(S^2,S^1) @>>> \mathbb{Z} @>>>0 \end{CD}$, from which one obtains $\pi_2(S^2,S^1) = \mathbb{Z}^2$.
\bibitem{note13} Pairwise linking numbers may distinguish the planar textures of the Hopf link, and certain others, but they certainly don't in the case of the Whitehead link ($H_1(\Sigma(L))=\mathbb{Z}_8$, 2 planar textures) or the Borromean rings ($H_1(\Sigma(L))=\mathbb{Z}_4\oplus\mathbb{Z}_4$, 4 planar textures). In these cases, for instance, we do not know the distinction.
\bibitem{note14} Note that the Stiefel-Whitney class of $\xi$ is the same as that of $L_\mathbf{n}$, one can observe that an orientation of $\xi$ is equivalent to a choice of normal vector for each 2-plane, which is just an orientation assigned to $\mathbf{n}$.
\bibitem{note15} One can see that this is true by constructing a section, $\hat{\mathbf{v}}$ of $\hat{\xi}$ explicitly on the 2-complex $X(\Omega^{w_1})$, where $\hat{\mathbf{n}} \big |_{X_1} \in \mathbb{R} \mathbb{P}^1$ and the winding around the boundary of each 2-cell is zero. If $\mathbf{v}$ is chosen to be non-zero on $X_1$, then by the Poincar\'{e}-Hopf theorem the number of zeros on each 2-cell is twice the degree of the extension of $\hat{\mathbf{n}}$ to that 2-cell, from which \eqref{eq:eulertwice} follows.
\end{thebibliography}
\end{document}